\documentclass{webofc}
\usepackage[varg]{txfonts}   
\usepackage{hyperref}

\begin{document}
\title{Data as a research infrastructure}
%
%
\subtitle{CDS, the Virtual Observatory, astronomy, and beyond}

\author{\firstname{Francoise} \lastname{Genova}\inst{1}\fnsep\thanks{\email{francoise.genova@astro.unistra.fr} ORCID iD 0000-0002-6318-5028} 
}
	
\institute{Université de Strasbourg, CNRS, Observatoire astronomique de Strasbourg, UMR 7550, F-67000 Strasbourg, France}

\abstract{%
  The situation of data sharing in astronomy is positioned in the current  general context of a political push towards, and rapid development of, scientific data sharing. Data is already one of the major infrastructures of astronomy, thanks to the data and service providers and to the International Virtual Observatory Alliance (IVOA). Other disciplines are moving on in the same direction. International organisations, in particular the Research Data Alliance (RDA), are developing building blocks and bridges to enable scientific data sharing across borders. The liaisons between RDA and astronomy, and RDA activities relevant to the librarian community, are discussed.
}
\maketitle
\section{Introduction}
\label{intro}
Astronomy has been a pioneer of scientific data sharing. Historical services such as the database of the International Ultraviolet Explorer, a UV satellite which was operational between 1978 and 1996 \cite{1995Ap&SS.228..385W}, or SIMBAD \cite{2000A&AS..143....9W}, operated by the CDS since 1983 succeeding the Bibliographic Star Index (BSI) and Catalogue of Stellar Identifications (CSI), were already accessible remotely before the advent of the web. For many years astronomers have been using remote data services in their daily research work: networked on-line services have been developed since the beginnings of the internet, and further federated in the Virtual Observatory (VO).

Data sharing thus became a routine for astronomers, and an established part of the research landscape for the discipline. But astronomy is not alone: science data sharing is currently a hot topic at the political level and for many funding agencies. Many disciplines are working at setting up data services and a disciplinary interoperability framework. Organisations such as the Research Data Alliance (RDA) work on building blocks and bridges to facilitate data sharing.
This paper describes several aspects of this fast evolving landscape, the political push towards Open Data (Section~\ref{sect-1}), the path towards and scientific impact of open data, taking astronomy as example (Section~\ref{sect-2}), examples of other disciplines for which data already is or is becoming a research infrastructure (Section~\ref{sect-3}), and the role of the Research Data Alliance (Section~\ref{sect-4}). The conclusion (Section~\ref{concl}) summarizes the fast evolution of the context and how astronomy should fit in, and the evolving role of librarians in data curation.

\section{The Open Data context}
\label{sect-1}
Scientific data sharing has been a hot topic in the recent period, including on the political agenda. The G8 Science Ministers made strong statements on the subject in June 2013\footnote{ \url{https://www.gov.uk/government/news/g8-science-ministers-statement}}, including:

\begin{itemize}
	\item "To the greatest extent and with the fewest constraints possible publicly funded scientific research data should be open"
	\item "Open scientific data should be easily discoverable, accessible, assessable, intelligible, useable, and wherever possible interoperable to specific quality standards"
\end{itemize}

They made additional statements in October 2015 \footnote{ \url{http://www.g8.utoronto.ca/science/2015-berlin.html}}, following a report of the Group of Senior Officials on Global Research Infrastructures (GSO), stating in particular that "further progress on sharing and managing scientific data and information should be achieved, especially by continuing engagement with community based activities such as the Research Data Alliance." Their Open Science statement of May 2016 \footnote{ \url{http://www.g8.utoronto.ca/science/2016-tsukuba.html} } proposes to establish a working group on open science, and to promote international coordination and collaboration to develop the necessary elements. 

Among the additional recent advances relevant to Open Data, one can cite the development of the FAIR Guiding Principles for scientific data management \cite{FAIR}. FAIR means Findable, Accessible, Interoperable, Reusable, and although this naming is recent and makes a buzz, the concepts have for a very long time been guiding principles for astronomy, where they are implemented and fully in use as discussed in Section~\ref{sect-2}. 

It is important to understand that scientific data sharing is not only a political subject and not only a way to enable the private sector to take the full benefit of scientific data, which can be the dominant objective of some of the Open Science promoters. A good description of the expected impact was given by the G8 Ministers of Science in their June 2013 statement: "We are committed to openness in scientific research data to speed up the progress of scientific discovery, create innovation, ensure that the results of scientific research are as widely available as practical, enable transparency in science and engage the public in the scientific process." 

Astronomy was a pioneer and can be taken as a case study of Open Science. In astronomy data is open, and moreover it is used by the scientific community in its daily research work. The way astronomers work now demonstrates that scientific data sharing first of all enables a change in paradigm in the way science is done. 

The layman view of astronomical infrastructures is, of course, our space and ground-based observatories. Data sharing is a powerful way to optimize the scientific return of these large infrastructures, as demonstrated by the well-known example of the bibliometric impact of the HST archive: more than half of the papers published each year from HST observations are using archival data\footnote{\url{https://archive.stsci.edu/hst/bibliography/pubstat.html}}. In addition, data sharing is indeed at the core of astronomy scientific needs, allowing multi-wavelength/multi-messenger/multi-technique astronomy, the study of time variability and of phenomena at different scales, the comparison of theoretical models with observations, etc. The fact that the community succeeded in setting up world-wide sharing of FAIR data, makes the distributed astronomical data one of the essential infrastructures of the discipline, as recognized for instance in the European Roadmap of astronomical infrastructures developed by ASTRONET\footnote{ASTRONET gathers the astronomy funding agencies operating in Europe} \cite{Roadmap2008}. As explained in Section~\ref{sect-2}, the observatory archives are not the only essential building blocks of this data infrastructure. For instance, the added-value services developed by the CDS, which are only one element of the landscape, receive more than 800,000 queries/day in average.
 
\section{The astronomical data infrastructure}
\label{sect-2}

\subsection{The basic elements of the astronomical data infrastructure}

The early definition and long term maintenance of a common data format, FITS \cite{1981A&AS...44..363W}, which includes both the data and relevant metadata, has certainly been one of the key elements to enable data sharing in astronomy. The discipline strong tradition of international collaboration, in particular to build telescopes and instruments, has also been a fertile ground for setting up the disciplinary data sharing framework beyond borders, together with the fact that observational data is in general open. Observations often become public after a proprietary period, which reserves data obtained from competitive Calls for Proposals to the original proposers for a limited period of time, an important factor to make data sharing acceptable by the community. Finally, the fact that the development of observatory archives and data services is driven by community needs has been essential to ensure widespread adoption of remote data usage as a routine of astronomers' scientific work.

The astronomical data infrastructure has very diverse components, among which of course is the observations: the archives from the ground and space based telescopes and the large sky surveys, which contain homogeneous observations of up to billions of objects. It also includes data from publications, value-added databases which gather homogenized information in particular from publications or observations, such as SIMBAD, VizieR \cite{2000A&AS..143...23O} and the Aladin image database \cite{BugaLISAVIII} at CDS, NED \cite{NEDLISAVIII}, and the ADS \cite{ADSLISAVIII} for bibliographic data, and more and more modelling results. 

SIMBAD, for instance, contains information gathered from catalogues and publications, including all the names of a given object. It is used  together with NED and VizieR, by observatory archives to transform object names provided by their users to their own user interface into coordinates, which are their main access key. On the other hand, the references in which the object was cited provided by SIMBAD are used, together with NED, by the ADS to provide the references citing an object. An additional added value is the capacity to sort the references attached to an object by relevance \cite{2015ASPC..492..284O}. It is important to note the huge amount of work behind the scenes, and the key role of the CDS "documentalists" \cite{2015ASPC..492...47B} \cite{COSIM}.

\subsection{Networking and Interoperability - The Astronomical Virtual Observatory}

This short description of SIMBAD already demonstrates that astronomical on-line services are networked. This networking begun as soon as the on-line resources began to be developed, in the early stages of the internet in 1993-1994. It was implemented through collaboration between the providers of added-value services, journals and observation archives, taking advantage of the capacity of the web to link web pages. It is still available and used, and is well illustrated for instance by the multiple links to distant resources available in the ADS. 

Around 2000, the idea to go beyond this already powerful networking of resources by providing seamless access to on-line data emerged as the Virtual Observatory concept. The Virtual Observatory is a framework of standards and tools which allows users to discover, access and use data. The interoperability standards are defined by the International Virtual Observatory Alliance \footnote{\url{http://ivoa.net}}.

The IVOA was created in 2002 as an alliance of national VO initiatives (it also includes the European VO initiative Euro-VO and the international organization ESA, (the European Space Agency). Its standardisation procedures are inspired from the World-Wide-Web Consortium (W3C). When possible, it uses generic elements, such as OAI-PMH (the reference for digital libraries), for its registry of resources \cite{2007ivoa.spec.0302H}, and W3C SKOS/RDF for its vocabularies \cite{2009ivoa.spec.1007G}. The VO standards deal with different aspects of the FAIR principles, in particular Findability, Accessibility and Interoperability. For instance, the IVOA registry of resources includes disciplinary extensions which allow users to discover resources of interest for them beyond the Dublin-Core-type criteria used in the library world.  

Like astronomy (compared for instance to particle physics), the Virtual Observatory has no central point; it stages a multipolar world in a global endeavour. It is an open and inclusive model, a thin interoperability layer on top of distributed and heterogeneous data holdings. Anyone can build a VO-enabled tool or register a data service. There are currently more than 100 "authorities" with a registered service, including large organisations as well as small research teams. The VO usage by data providers has been evolving beyond enabling interoperable access to their data, and more and more of them now embed VO building blocks in their archives and services.

\subsection{Current status of the Virtual Observatory}

The VO is currently operational and used. Its successful development and implementation has three pillars, as recognized from the early phase of the European VO coordination \cite{2006ASPC..351..771P}: support to science users in their uptake of the VO, support to data providers in their uptake of the VO, and the technological work to develop and update the VO standards and tools. The VO has several kinds of users:  scientists using VO-enabled data services and tools, data providers implementing the VO framework, and developers of VO-enabled tools. The VO is, most of the time, invisible from its science users - the price to pay when one really aims to provide seamless access. One key for success amongst science users has been the high quality of data and services available in the VO. This is thanks to the work of the data and service providers. Another one is seamless access to data through interoperable tools such as Aladin or Topcat, which are well suited to science needs thanks to the work of the VO developers. The VO was adopted by the data and service providers because it gives more visibility to their data by making them discoverable, and also because it provides access tools and building blocks for their systems: there is no need to reinvent the wheel, the VO developers already worked and pooled resources and expertise to propose relevant and handy solutions.

The current priorities of the IVOA are linked to the needs of the future large astronomical projects: multi-dimensional data, with a first milestone reached in May 2017 with a complete initial set of relevant standards; and time domain. The current European VO project is closely aligned with these priorities. The European Commission has been funding a series of projects since 2001 to support the coordination of VO activities in Europe \cite{2015A&C....11..181G}. The current one is a Work Package of the \emph{Astronomy ESFRI\footnote{ESFRI \url{http://www.esfri.eu/} is the European Strategic Forum for Research Infrastructures, which develops a roadmap for pan-European Research Infrastructures} and Research Infrastructure Cluster} ASTERICS\footnote{\url{https://www.asterics2020.eu/} , 2015-2019},  called \emph{Data Access, Discovery and Interoperability} (DADI), which is foreseen for 4.5 million euros over 4 years starting in May 2015 (the whole budget of ASTERICS is 15 million euros). DADI aims at "making the ESFRI and pathfinder project data available for discovery and usage by the whole astronomical community, interoperable in a homogeneous international framework (the Virtual Observatory), and accessible by a set of common tools." DADI and ASTERICS involve both astronomy and astroparticle physics telescopes, which is particularly important with the rapid development of multi-messenger astronomy. DADI includes the Euro-VO partners, i.e. VO initiatives from France, Germany, Italy, Spain and UK, and representatives of the large projects Cherenkov Telescope Array (CTA), the European Gravitational Observatory (EGO)/Einstein Telescope (ET), the Cubic Kilometre Neutrino Telescope (KM3Net) and the Square Kilometre array (SKA) and their pathfinders. The European Southern Observatory (ESO), which is in charge of the Extremely Large Telescope (ELT), is associated to the project, which is also working in close collaboration with ESA.

\subsection{Big and smaller data in astronomy}

Another trendy word in the data world is "Big Data." Astronomy can be taken as an example of "Big Data" science, because the ground and space-based telescopes and the large surveys produce large data volumes. But the discipline also takes care of its "smaller data", the so-called Long Tail of data \cite{Borgman}, in particular the research results which are linked to publications. This started in 1993, with the agreement between the CDS and \emph{Astronomy \& Astrophysics} that the "long" tables in the papers would be published on line in an ftp server by the CDS \cite{1995VA.....39..227O}. The agreement was later extended to other journals. These data come with a standard description, which has been the basis to ingest them in the VizieR system when it was implemented. VizieR can store any kind of astronomical quantity which can be found in a publication. One can note here that variety is also one of the characteristics of Big Data. The standard description and the UCD controlled vocabulary for astronomical quantities \cite{2007ivoa.spec.0402M} provide a homogeneous frame for the very heterogeneous database content. In August 2017 VizieR contains more than 16,000 "catalogues", which are actually data sets which can include tabular as well as non-tabular data such as images, spectra, time series, etc. attached to a publication, or object lists and other data from large surveys \cite{Ocvirk}.

VizieR approach to the Long Tail of Data is thus to preserve data validated by a publication, and to make it fully discoverable and interoperable in the Virtual Observatory, usable and citable through ADS (and soon through a DOI). "VizieR Photometry Viewer" \cite{2014ASPC..485..219A} is an example of value-added service made possible by the homogeneous data description and the rich information stored about the data \cite{Landais}: it extracts all the spectral points available in VizieR in a given direction on the sky, and displays them on a single plot taking into account the photometric system used in the original observation. Each point is linked to the original reference.

In astronomy there is thus not really "big" and "little" data, but rather Useful, Validated and Documented data which are fully FAIR.

\section{Data as an infrastructure: examples from different disciplines}
\label{sect-3}

The notion and management of Research Infrastructures can be different from country to country. This is a domain where European coordination made a difference and enabled the development of a regional strategy, which includes global infrastructures such as SKA. The European landscape of Research Infrastructures is structured by the ESFRI Roadmap, which covers all scientific disciplines and is regularly updated. The ESFRI Research Infrastructures are infrastructures of pan-European interest defined as "facilities, resources or services of a unique nature identified by European research communities to conduct top-level research activities in all fields". The first Roadmap was published in 2006, the last update \cite{ESFRI2016} in 2016, and the next version is foreseen for 2018. More and more European countries also develop a National Roadmap, taking into account the ESFRI Roadmap as well as their national priorities. 

All research infrastructures have aspects linked to data, and ESFRI Roadmap candidates have to answer questions about their data management and policy, and their computing and network needs. This is also the case for instance for the French National Roadmap candidates. Some of the infrastructures on the Roadmaps are fully "virtual", for instance the CDS is in the French Roadmap. 

The way different disciplines develop their data infrastructure is strongly dependent from the disciplinary culture and organisation. It is interesting to observe that the Humanities have been present in the ESFRI Roadmap from the beginning, with two projects, DARIAH and CLARIN, which are distributed data infrastructures: DARIAH\footnote{http://www.dariah.eu/} supports digital research in the arts and humanities in general, CLARIN\footnote{https://www.clarin.eu/} is focused on digital language resources. In France, the national Humanities research infrastructure Huma-Num\footnote{http://www.huma-num.fr/about-us} provides added-value services in support of the data life cycle and in addition manages disciplinary consortia set up for 4 years (with possible renewal) to organise communities around data sharing topics relevant to their needs. This is very useful for disciplines which are more loosely organised than astronomy. It also coordinates French participation in DARIAH and CLARIN. The national organisation in Earth Sciences is different: there are 4 disciplinary "Poles": Solid Earth, Ocean, Atmosphere, Continental Surfaces and Interfaces, in which all the relevant research organisations, including the French Space Agency CNES, participate. "Inter-Poles" Technical discussions and collaboration on topics of common interest are organised. An overarching structure is being built, which is a candidate for the national Roadmap. Each Pole has strong links with the ESFRI and international projects in its domains.

The way different disciplines organise themselves to tackle the development of their disciplinary interoperability framework was discussed during a session  \cite{Session44} of the SciDataCon 2016 Conference\footnote{http://www.scidatacon.org/2016/}, which was held in Denver (USA) on 2016, September $11^{th}-13^{th}$. The panel included representatives from humanities, linguistics, astronomy, earth sciences, material sciences and crystallography. Many commonalities were found: 
\begin{itemize}
	\item the development of the data infrastructure must be science driven; 
	\item defining the disciplinary part of the interoperability standards is mandatory but difficult; 
	\item it is important not only to share data, but also tools, user interfaces, etc; 
	\item sociological aspects are more challenging than the technical ones. 
\end{itemize}

Once again, incentives to data sharing appeared to be a key issue for all. Governance is more diverse, linked, as explained earlier, to the discipline organisation and history. It was also found that many sociological and technological aspects can be shared, and it was suggested that the RDA is a good vehicle for that, with roles also for  CODATA\footnote{Committee on Data for Science and Technology, \url{http://www.codata.org/} } and WDS\footnote{World Data System, \url{http://www.icsu-wds.org/} } of the International Council for Science ICSU. 

As explained in \cite{Session44}, international organizations such as GEO, CODATA and the RDA can be used to help disciplines which are not strongly organised at the international level to discuss their data framework. International discussion forums can be established as CODATA Task Groups and/or RDA Interest and Working Groups (Section~\ref{sect-4}). In addition, the WDS “community of excellence” is a host for individual data providers as regular members, but also for disciplinary networks once they are constituted, and it organizes interaction between its members. The IVOA is a Network member of the WDS, and several astronomical data centres, including the CDS, are regular members.

\section{The Research Data Alliance}
\label{sect-4}

The Research Data Alliance\footnote{https://www.rd-alliance.org/} is a community-driven international organisation created in 2013 by the Australian government, the European Commission and the USA NSF, with the mission to build the social and technical bridges to enable open sharing of data. In August 2017, the RDA has more than 5900 members from 129 countries, 66\% from academia and research, 14\% from public administration, 11\% from enterprise and industries. It hosts focused Working Groups and Interest Groups tackling many different aspects of data interoperability challenges, in themes such as domain science, community needs, data referencing and sharing, data stewardship and services, or base infrastructure. Any member can propose a Group, and in August 2017 there are 88 active Groups, 30 Working Groups which prepare implementable deliverables in 18 months, and 58 Interest Groups which organise discussion and exchange of good practices on a specific topic and can also produce outputs. The activities of the Working Groups and Interest Groups, including their email exchanges and the documents they produce, are documented on the RDA web site, as well as the recommendations and outputs, and adoption cases. RDA members have a large variety of profiles, which makes it a remarkable international, neutral forum gathering very diverse expertise. 
They include data providers, researchers, project and program managers, and publishers. Librarians make more than 10\% of membership. 

There is not Astronomy Group in the RDA, because it would duplicate the IVOA, but astronomy is present, and the experience gained in building the IVOA and the astronomical data infrastructure is used in the RDA context. For instance, the author of this paper is currently co-chair of the RDA Technical Advisory Board. It is also interesting to note that the RDA Group working on an International Materials Resource Registry will reuse the concepts of the IVOA Registry of Resources. The IVOA Registry of Resources is already customized to their needs by planetary sciences and the Virtual Atomic and Molecular Data Centre VAMDC. The RDA allows wider dissemination. Data experts from the astronomical community participate in several other RDA Groups, for instance those dealing with Data Provenance. Astronomy is also involved in the \emph{Disciplinary Collaboration Framework Interest Group} recently proposed as an inter-disciplinary forum for discussion, which aims at becoming the voice of scientific disciplines in the RDA.   

Many RDA activities are of interest for the LISA community. In particular, one of the RDA recommendations is a catalogue of requirements \cite{certcriteria} and a set of procedures \cite{certprocedure} for core audit and certification of data repositories. It was prepared by the \emph{Repository Audit and Certification DSA–WDS Partnership Working Group}, which gathered members from the Data Seal of Approval DSA\footnote{\url{http://www.datasealofapproval.org}} and the World Data System. These two organisations had been managing two different basic certification processes for data repositories, which were merged and improved by the Working Group. This is a very important clarification and simplification of the landscape, at a time when it is more and more important for data providers to demonstrate that there are trustworthy through certification. The CDS for instance had successively and successfully applied for WDS and then DSA certification \cite{Landais} because it was not easy to choose one. Now it is looking for renewal of its certification in the unified system.

Another remarkable aspect of the RDA is the importance given to data publishing and persistent identifiers. For instance, a Working Group made widely adopted recommendations for dynamic data citation (when data evolves in time keep track of the queries in a "query store"). Several Working Groups involving publishers and data providers, managed in collaboration between the RDA and the WDS, have also been tackling different aspects of data publishing. This includes for instance Publishing Data Workflows and Bibliometrics, and  currently the \emph{Scholarly Link Exchange Working Group} (Scholix) which aims to enable a comprehensive global view of the links between scholarly literature and data.

The exchange of good practices and lessons learnt in the \emph{Interest Group on the Long Tail of Research Data} and the one on Domain Repositories can also be of interest for the LISA community. Finally, the \emph{Libraries for Research Data Interest Group} produced a very successful output, \emph{23 Things: Libraries for Research Data} \cite{23Things}, which provides an overview of practical, free, online resources and tools that one can begin using today to incorporate research data management into one's practice of librarianship. The document is available in 11 languages\footnote{https://www.rd-alliance.org/group/libraries-research-data-ig/outcomes/23-things-libraries-research-data-supporting-output}.

\section{Conclusions}
\label{concl}

These are exciting times for scientific data sharing! Astronomy has been at the forefront and, thanks to the efforts of many organisations and individuals, data is now one of the disciplinary research infrastructures. This includes many different distributed on-line resources, the IVOA being a rare example of an operational, global interoperability framework. Even if many astronomers do not realize it, the Virtual Observatory enables scientists to routinely Find, Access and Interoperate data and tools. 

The context of scientific data sharing is evolving very fast, and many disciplines are moving on. It is important for the astronomy community to be involved in initiatives such as the RDA, to ensure that our requirements are taken into account in the emerging data sharing building blocks. Trust is an essential component of the landscape. It mainly relies on the quality of data and of the accompanying documentation, which requires lots of underlying work at the disciplinary level by people who know the data. The astronomy data services have been building trustworthiness with their user community for a long time (45 years for the CDS). In the new context, it can be worth applying for an external certification. Even if the process is not completed up to submission of an application for review the catalogue of criteria is a very good basis for in-house assessment of the repository processes.

The evolving role of librarians has been an important topic of the LISA conferences for a while. Specialized librarians, the so-called "documentalists", have been one of the essential profiles at CDS since the beginning, one of its three pillars with researchers and IT engineers \cite{2015ASPC..492...13P}. There are several levels of data curation, including disciplinary repositories, non-specialist repositories such as Zenodo, Figshare or the institutional ones set up by many Universities or research organisations. Going beyond Dublin-Core-type data description requires disciplinary knowledge, and discipline-savvy librarians have a key role to play in the new research landscape, now and in the future.

\section{Aknowledgements}

The author acknowledges support from the Research Data Alliance - Europe 3 and ASTERICS Projects, funded by the European Commission (projects 653194 and 653477 respectively).

\bibliography{mybibfile}

\end{document}